# Raman Signature and Phonon Dispersion of Atomically Thin Boron Nitride

*Qiran Cai,[1] Declan Scullion,[2] Aleksey Falin,[1] Kenji Watanabe,[3] Takashi Taniguchi,[3] Ying Chen,[1]\* Elton J. G. Santos[2,4]\* and Lu Hua Li[1]\**

1. Institute for Frontier Materials, Deakin University, Geelong Waurn Ponds Campus, Victoria 3216, Australia

2. School of Mathematics and Physics, Queen's University Belfast, Belfast BT7 1NN, United Kingdom

3. National Institute for Materials Science, Namiki 1-1, Tsukuba, Ibaraki 305-0044, Japan

4. School of Chemistry and Chemical Engineering, Queen's University Belfast, BT9 5AL, United Kingdom.

AUTHOR INFORMATION

**Corresponding Author**

*Email: ian.chen@deakin.edu.au; e.santos@qub.ac.uk; luhua.li@deakin.edu.au

ABSTRACT: Raman spectroscopy has become an essential technique to characterize and investigate graphene and many other two-dimensional materials. However, there still lacks consensus on the Raman signature and phonon dispersion of atomically thin boron nitride (BN), which has many unique properties distinct from graphene. Such a knowledge gap greatly affects the understanding of basic physical and chemical properties of atomically thin BN as well as the use of Raman spectroscopy to study these nanomaterials. Here, we use both experiment and simulation to reveal the intrinsic Raman signature of monolayer and few-layer BN. We find experimentally that atomically thin BN without interaction with substrate has a G band frequency similar to that of bulk hexagonal BN, but strain induced by substrate can cause pronounced Raman shifts. This is in excellent agreement with our first-principles density functional theory (DFT) calculations at two levels of theory, including van der Waals dispersion forces (opt-vdW) and a fractional of the exact exchange from Hartree-Fock (HF) theory through hybrid HSE06 functional. Both calculations demonstrate that the intrinsic $E_{2g}$ mode of BN does not depend sensibly on the number of layers. Our simulations also suggest the importance of the exact exchange mixing parameter in calculating the vibrational modes in BN, as it determines the fraction of HF exchange included in the DFT calculations.

**TOC GRAPHICS**

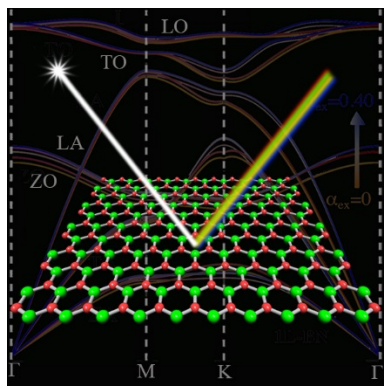

**KEYWORDS** boron nitride, 2D nanomaterials, Raman spectroscopy, phonon dispersion, strain control

Raman spectroscopy has become an essential technique to characterize and investigate graphene and many other two-dimensional materials. However, there still lacks consensus on the Raman signature and phonon dispersion of atomically thin boron nitride (BN) which has many unique properties distinct from graphene. Such a knowledge gap greatly affects the understanding of basic physical and chemical properties of atomically thin BN as well as the use of Raman spectroscopy to study these nanomaterials. Here, we use both experiment and simulation to reveal the intrinsic Raman signature of monolayer and few-layer BN. We find experimentally that atomically thin BN without interaction with substrate has a G band frequency similar to that of bulk hexagonal BN, but strain induced by substrate can cause pronounced Raman shift. This is in excellent agreement with our first-principles density functional theory (DFT) calculations at two levels of theory, including van der Waals dispersion forces (opt-vdW) and a fractional of the exact exchange from Hartree-Fock (HF) theory through hybrid HSE06 functional. Both calculations demonstrate that the intrinsic $E_{2g}$ mode of BN does not depend sensibly on the number of layers. Our simulations also suggest the importance of the exact exchange mixing parameter in calculating the vibrational modes in BN, as it determines the fraction of HF exchange included in the DFT calculations.

Boron nitride (BN) nanosheets, atomically thin hexagonal boron nitride (hBN), are an isoelectronic and structural analog of graphene, with excellent mechanical strength and thermal conductivity.[1,2] BN nanosheets have a number of properties and applications distinct from graphene.[3] For instance, BN nanosheets can withstand oxidation beyond 800 °C in air,[4] and this superior thermal stability makes them an excellent barrier to protect metals against oxidation at high temperatures.[5,6] Furthermore, BN nanosheets with wide band gaps of ~6.0 eV are the thinnest electrically insulating materials,[7,8] suitable to serve as dielectric substrates to improve the mobility of graphene and molybdenum disulfide ($MoS_2$) based devices.[9,10] BN nanosheets can also enable highly sensitive, reproducible and reusable sensors.[11-15]

Raman spectroscopy is an indispensable method to characterize and study graphene and many other 2D nanomaterials, and tremendous efforts have been devoted to this field. Take graphene as an example. The thickness of single and few-layer graphene can be unambiguously determined by the Raman intensity ratio between its G and 2D bands, and the detailed structure of the 2D band.[16-18] Similarly, $MoS_2$, a typical transition metal dichalcogenide, shows clear Raman frequency changes in $E_{2g}^1$ and $A_{1g}$ modes when scaled down to atomically thin sheets.[19] In addition, Raman can be used to probe crystallinity,[20-22] edge state,[23-25] strain,[26-33] doping,[34-39] lattice temperature[40-42] of these 2D nanomaterials. In spite of the similar structure to graphene, BN nanosheets do not show a Raman 2D band due to the lack of Kohn anomaly, but Raman spectroscopy is still widely used to characterize BN nanosheets.[4,8,10,12,43-52] Therefore, it is important to study Raman signature and phonon dispersion of BN nanosheets of different thicknesses.

However, there still lacks consensus on the Raman signature of atomically thin BN. Gorbachev et al. first reported that the Raman G band of atomically thin BN shifted with thickness: compared to bulk hBN crystals, monolayer (1L) BN on silicon oxide covered silicon wafer (SiO₂/Si) substrate showed upshifted G band, while downshifts were observed from few-layer BN on the same substrate.[44] Since then, these results have been used as a reference to determine the thickness of BN nanosheets in many reports. However, we found[4,8,12] that 1L and few-layer BN nanosheets mechanically exfoliated from hBN single crystals from the same source on SiO₂/Si substrate all showed upshifted G bands, but to different extent, compared to bulk hBN. The discrepancy implies a knowledge gap and can greatly affect the use of Raman spectroscopy for characterizing BN nanosheets. Till now, there is still no systematic study to understand the intrinsic Raman signatures of atomically thin BN.

Here, we used both experiments and theoretical calculations to reveal the phonon dispersion and Raman signatures of BN nanosheets of different thicknesses. We found that the intrinsic Raman frequency of BN did not show dramatic change with thickness, but strain induced by substrates could greatly affect the G band frequency of monolayer and few-layer BN. This study not only provides a fundamental understanding of the vibrational property of atomically thin BN but also guides the use of Raman spectroscopy to analyze these 2D nanomaterials.

The high-quality atomically thin BN crystals used in this study were mechanically exfoliated from hBN single crystals[53,54] on SiO₂/Si substrates with and without pre-fabricated wells (1.3 μm in diameter and >2 μm deep) by Scotch tape technique. Our previous publications can be referred for details on the exfoliation process.[4,8,12,15] Atomically thin BN nanosheets were located and identified using an optical microscope (Figure 1a and d) thanks to the interference enhancement by the SiO₂ layer on Si wafer.[44] The thickness of the BN nanosheets was then measured by atomic force microscopy (AFM). Figure 1b and e show AFM images of 1-2L BN nanosheets bound to SiO₂/Si substrate and suspended over the wells, respectively. According to the corresponding height traces, the thicknesses of 1L and 2L BN were about 0.5 and 0.9 nm, respectively. These results are consistent with previous reports.[6,8,12,44]

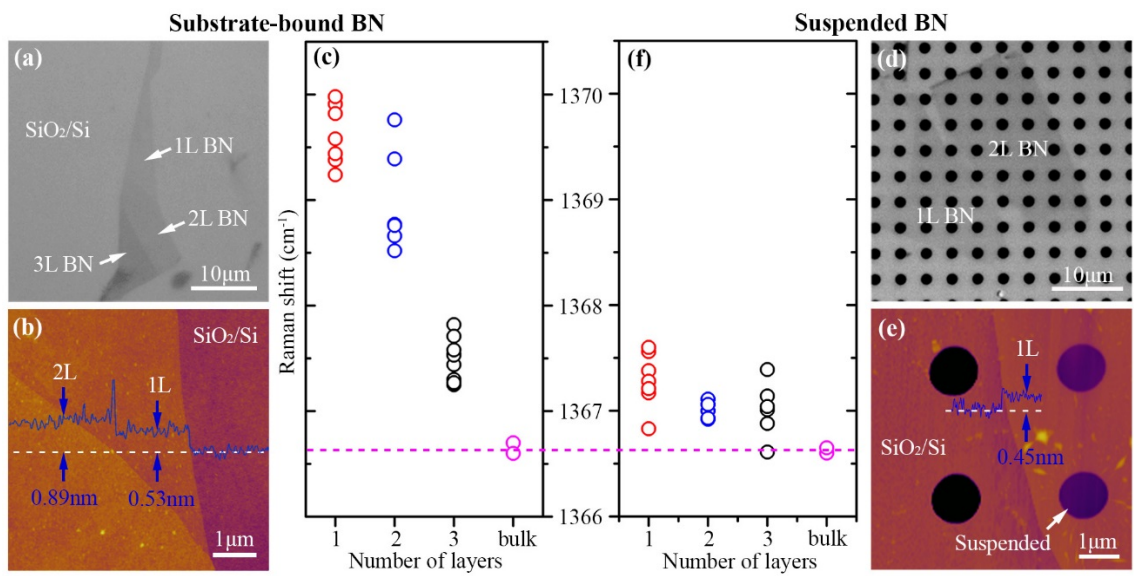

**Figure 1.** (a) Optical image of 1-3L BN nanosheets bound to SiO₂/Si substrate; (b) the corresponding AFM image with height trace inserted; (c) Raman frequencies of the G band of substrate-bound 1-3L and bulk BN; (d) optical image of 1-2L BN nanosheets partially suspended

over ~1.3 μm wells; (e) the corresponding AFM image with height trace inserted; (f) Raman frequencies of the G band of suspended 1-3L BN and bulk hBN.

The G band frequencies of substrate-bound 1-3L BN in comparison to that of bulk hBN crystal are shown in Figure 1c. Bulk hBN crystals showed a G band centered at 1366.6±0.2 cm$^{-1}$ (based on 6 samples, *i.e.* $N$=6), consistent with previous studies.[4] It corresponds to the $E_{2g}$ vibration mode in hBN.[55,56] Obviously upshifted G band were observed on the substrate-bound atomically thin BN: 1369.6±0.6 cm$^{-1}$ for 1L ($N$=8), 1369.0±0.5 cm$^{-1}$ for 2L ($N$=6), and 1367.5±0.2 cm$^{-1}$ for 3L ($N$=9). In addition, 1-3L BN showed more variations in their G band frequency than the bulk crystals.

Three reasons may cause the upshifted G band of atomically thin BN: 1) doping due to substrate and/or adsorbates; 2) heating effect from laser beam; 3) strain induced by substrate (*i.e.* roughness). The doping effect can be ruled out because previous studies showed that monolayer BN nanosheets were not subject to surface doping.[39,57] This is understandable by considering their electrical insulation. The laser heating effect can also be eliminated because we found that the G band frequency of BN nanosheets downshifted at higher temperatures: the G band of a 1L BN downshifted to 1367.7 cm$^{-1}$ at 75 °C and 1366.4 cm$^{-1}$ at 100 °C, respectively. In other words, laser heating should increase the temperature and soften the $E_{2g}$ phonons, resulting in redshift of the G band.[56,58] Thus, laser heating cannot explain the upshifts of substrate-bound atomically thin BN. This leaves us with the only factor: strain.

Atomically thin nanosheets which have small bending moduli[59,60] tend to corrugate or ripple, and it has been found that the uneven surface of $SiO_2$ introduces strain to graphene.[61,62] Therefore, it is likely that the different Raman frequency shifts from BN nanosheets of different thicknesses are due to their different degrees of corrugation and hence strain on $SiO_2$/Si substrate. In other words, suspended atomically thin BN nanosheets which are mostly free from substrate disturbance (though probably still not completely strain free) should be much better to illustrate their intrinsic Raman signatures. As shown in Figure 1f, the average G band frequencies of suspended BN nanosheets were very close: 1367.3±0.3 cm$^{-1}$ for 1L, 1367.0±0.1 cm$^{-1}$ for 2L, and 1367.0±0.2 cm$^{-1}$ for 3L, respectively. These values were very close to the Raman frequency of bulk hBN (*i.e.* 1366.6±0.2 cm$^{-1}$). As aforementioned that BN nanosheets are not subject to substrate doping, the only difference between the substrate-bound and suspended BN nanosheets should be strain. Our results imply that 1) the observed different G band frequency of 1-3L BN on $SiO_2$/Si should be due to their different flexibility and levels of strain caused by the uneven substrate; 2) the intrinsic Raman frequency of atomically thin BN of different thicknesses may be close to each other and that of bulk hBN crystals.

To better understand the vibrational properties of BN nanosheets, we performed first-principles density functional calculations including van der Waals (vdW) dispersion forces (see *Experimental section*). We also took into account a fractional component of the exact exchange from the Hartree-Fock (HF) theory hybridized with the density functional theory (DFT) exchange-correlation functional at the level of the range-separated HSE06 hybrid functional. The exchange-correlation energy in HSE06 is given by:

$$E^{HSE}{}_{XC} = \alpha_{ex} E_X{}^{HF,SR}(\omega) + (1 - \alpha_{ex}) E_X{}^{\omega PBE,SR}(\omega) + E_X{}^{\omega PBE,LR}(\omega) + E_C{}^{PBE} \qquad (1)$$

where $E_X{}^{HF,SR}(\omega)$ is the short-range (SR) HF exchange; $E_X{}^{\omega PBE,SR}(\omega)$ and $E_X{}^{\omega PBE,LR}(\omega)$ are the short and long range (LR) components of the PBE exchange functional, respectively; $\omega = 0.20\,\text{Å}^{-1}$ is the screening parameter, which defines the separation of the SR and LR exchange energy; and $\alpha_{ex}$ is the HF mixing factor that controls the amount of exact Fock exchange energy in the functional.[63-65] Note that HSE functional with $\alpha_{ex} = 0$ becomes the PBE functional.[66] Therefore, any limitation of the exchange and correlation functional in the chemical and physical description of the vibrational modes could be improved.

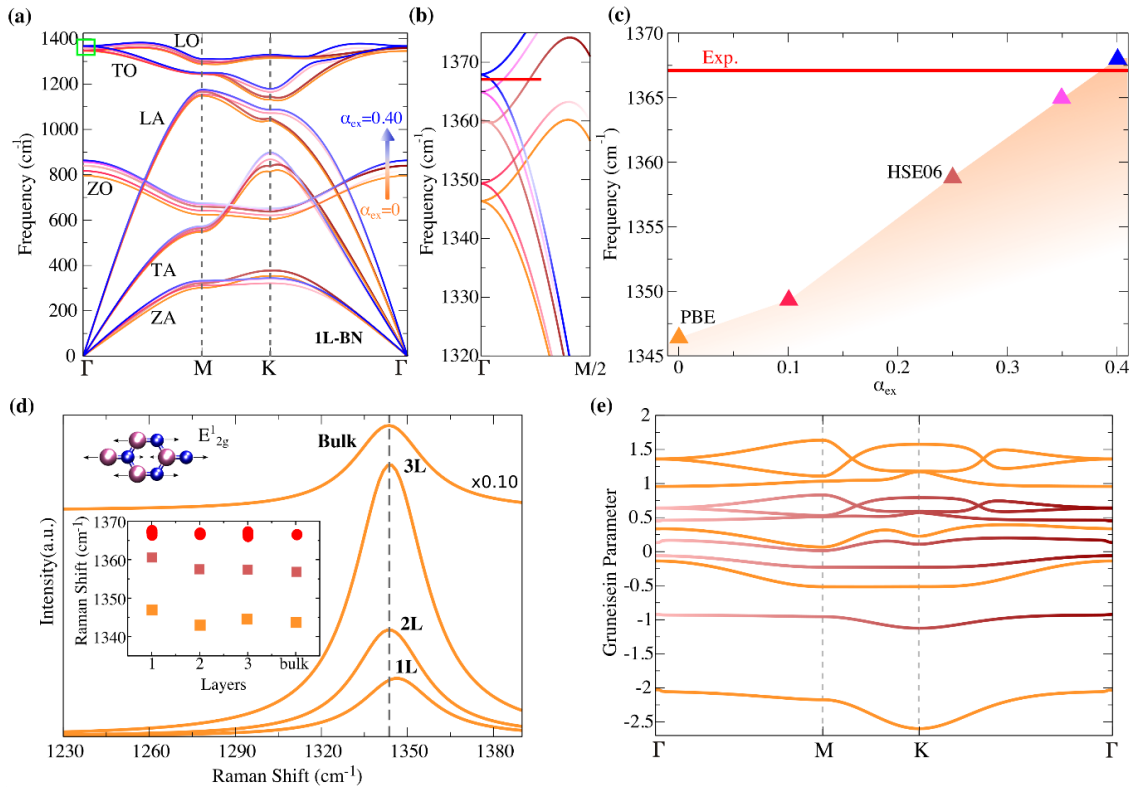

**Figure 2.** (a) Phonon dispersion with their acoustic (ZA, TA, LA) and optical (ZO, TO, LO) branches labeled for 1L BN nanosheet calculated within the density functional theory. The range-

separated hybrid functional HSE06 was used at different values of the mixing factor $\alpha_{ex}$. The frequency of phonon branches increases with increased $\alpha_{ex}$, as displayed in different colors. (b) The enlarged region along the optical mode ($E_{2g}$) highlighted by a small green rectangle in (a) to compare HSE06 simulations with the experimental value (red bar). (c) Calculated Raman shifts (cm$^{-1}$) for $E_{2g}$ mode as a function of $\alpha_{ex}$ for 1L BN system. The different values of $\alpha_{ex}$ correspond to different contribution of HF exact exchange to the exchange-correlation functional: 0.0 (PBE including vdW corrections, opt-vdW functional), 0.10, 0.25 (standard HSE06), 0.35, and 0.40. (d) *Ab initio* Raman spectra of free-standing $N$L ($N$=1-3), with the vertical dashed line representing the calculated $E_{2g}$ frequency of bulk hBN calculated at opt-vdW level. The top inset shows a schematic of the Raman-active mode $E_{2g}$ in BN, and the bottom inset shows the variation of the Raman shift as a function of thickness at standard HSE06 functional (faint brown squares) and opt-vdW (orange squares) in comparison with the experimental values (red circles). (d) Phonon Grüneisen dispersion relations calculated for 1L BN layer at standard HSE06 (faint brown) and opt-vdW (orange) levels of theory.

Figure 2a shows the *ab initio* phonon dispersion for 1L BN at different values of $\alpha_{ex}$ using the HSE06 method. Although the different values of $\alpha_{ex}$ gave rise to similar trends in the dispersion of the phonon modes along the Brillouin-zone, including acoustic (ZA, TA, LA) and optical (ZO, TO, LO) phonon modes, the phonon frequencies slightly increased with increased $\alpha_{ex}$. The largest difference in frequency from PBE ($\alpha_{ex} = 0$) and HSE06 ($\alpha_{ex} = 0.40$) methods was 47.80-67.41 cm$^{-1}$ at the ZO phonon branch; smaller differences were observed at specific **k** points, such as the $\Gamma$ point (Figure 2b). In terms of the optical LO and TO phonons which are responsible for the Raman G band of BN, PBE and standard HSE06 ($\alpha_{ex} = 0.25$) gave rise to ~1.5% and 0.5% underestimated frequencies than the experimental values from the suspended 1L samples (*i.e.* 1367.3±0.3 cm⁻¹), respectively. However, this difference became negligible once higher magnitudes of $\alpha_{ex}$ ($= 0.40$) were included (Figure 2c). This indicates that even though the exact HF-exchange at its default mixing value corrects the self-interaction error in DFT and gives a better description of the vibrational properties,[67,68] a full representation of the experimental data is only possible at higher values of $\alpha_{ex}$. This highlights the importance of the HF exact exchange in the LR asymptotic behavior of the exchange-correlation potential, $-1/r$. Although it has been shown previously that $\alpha_{ex}$ could improve the calculations of the Raman frequencies in several non-layered systems,[69] no attempt has been made to improve the description of the vibrational properties of BN using approaches beyond mean-field level or quantum chemistry based methods. In fact, most of the discrepancy between calculations and experiments have been observed at local or semi-local levels using different functionals.[70-78] Our simulations at HSE06 level gave rise to vibrational values much closer to those from experiments.

Figure 2d shows the *ab initio* Raman spectra of freestanding 1-3L BN nanosheets and bulk hBN using a 514.5 nm laser simulated by PBE (opt-vdW functional) (main panel). The polarization of the incident and scattered light was set along (0, 1, 0) plane. Based on the opt-vdW functional, the in-plane vibrational $E_{2g}$ mode of freestanding 1L, 2L, 3L and bulk BN nanosheets were at 1348.5, 1343.3, 1347.6 and 1343.7 cm⁻¹, respectively (top inset in Fig. 2d). These frequencies were clearly downshifted relative to our experimental values (circles in red). In comparison, improvement was achieved at the level of HSE06 (faint brown squares) using the standard mixing value of $\alpha_{ex} = 0.25$

: 1360.6, 1357.6, 1357.4 and 1356.8 cm$^{-1}$ for 1L, 2L, 3L and bulk hBN, respectively (bottom inset in Fig. 2d). Although the difference between calculations and experiments could be further corrected using higher values of $\alpha_{ex}$, both sets of simulations reproduced closely the trend observed in the experiments (Figure 1f). That is, the $E_{2g}$ mode did not depend sensibly on the number of BN layers. This behavior is different from many other 2D nanomaterials, whose Raman modes change noticeably with decreased number of layers.[16-19]

Based on the phonon-dispersion relations, we calculated the Grüneisen parameters ($\gamma$) at different **k** points of the Brillouin zone (see **Experimental section**) at opt-vdW and HSE06 ($\alpha_{ex} = 0.25$) levels (Figure 2e). HSE06 gave smaller Grüneisen parameters than opt-vdW. As previously discussed (Figure 2a), this was due to the effect of the increment of the phonon frequencies, as some fractional parts of the exact exchange were taken into account. Most of the HSE06-deduced values lay in the range of $-1.13$ to $+0.83$ at **K** and **M** points, respectively, and can be grouped into two sets: i) modes with $\gamma > 0$, which were formed by LO, TO, LA and TA phonon branches; and ii) modes with $\gamma < 0$, where ZO and ZA were the main components. The Grüneisen parameter for the LO and TO modes at $\Gamma$ was 0.64. As shown later, this value is useful for the calculation of deformation or strain in BN from its *G* band shift. It should also be mentioned that the phonon frequencies were very sensitive to strain in the utilized supercell: changes as small as ~0.15% in lattice constant used in our calculations could induce ~9 cm$^{-1}$ in Raman shift in both opt-vdW and HSE06 simulations. Such Raman shift increased or decreased linearly with further applied strain, suggesting a strain-driven vibrational state in BN nanosheets.

To further show the effect of strain on the Raman spectrum of BN nanosheets, we experimentally modified the level of strain in atomically thin BN on $SiO_2$/Si and studied their G band frequency. Due to the different thermal expansion coefficient (TEC) between BN and $SiO_2$, heat treatment can further increase the corrugation and hence strain in atomically thin BN, similar to the case of graphene.[79] In our experiment, atomically thin BN nanosheets were heated up to 400 °C in argon (Ar) gas for 1 h. The roughness of a 1L BN before and after heat treatment was revealed by AFM (Figure 3a and b). The AFM images show that the heat treatment did increase the roughness of the nanosheet. Such change can be better seen from the Gaussian fitted height distributions in Figure 3c: a relatively broader height distribution after heat treatment, corresponding to approximately 20% of increased roughness. Height-height correlation function (HHCF) is another direct indication of the total and short-range roughness. The function can be described as:

$$g(x) = \langle (h(\vec{x}) - h(\vec{x} - \vec{r})^2 \rangle \tag{2}$$

where $\vec{x}$ is any specific point in the image, and $\vec{r}$ is a displacement vector. The average height difference between any two points separated by the distance $r$ is described by the function $g(x)$.[80] For the self-affine scaling, the function can be simplified as:

$$g(x) = Ar^{2H} \text{ for } r \ll \xi \tag{3}$$

$$g(x) = 2\sigma^2 \text{ for } r \gg \xi \tag{4}$$

where A is a constant; $H$ describes the degree of surface irregularity (*i.e.* jag) at the short range; $\xi$ is the correlation length; $\sigma$ denotes the root mean square (RMS) roughness.[81] Figure 3d shows the log-log plot of the HHCF and the corresponding fittings (Eq. 3 and 4) of the 1L BN before and after heat treatment. The correlation functions increased significantly in the short range (*i.e.* r< $\xi$), and the fitting deduced *2H* values for the 1L BN before and after heat treatment were 1.75±0.05 and 1.51±0.05, respectively. This strongly suggests a more corrugated surface after annealing. In addition, the RMS roughness amplitude σ of the 1L BN increased from 123.6±0.1 pm to 139.4±0.4 pm after heat treatment. According to the intersection of power-law line and the saturation line, the correlation length ξ can be calculated by:

$$\xi = (2\sigma^2/A)^{1/2H} \tag{5}$$

The values of ξ for the 1L BN before and after heat treatment were 25.7±0.1 and 24.2±0.2 nm, respectively. The larger RMS roughness and smaller correlation length further indicate larger roughness in the 1L BN after annealing. Therefore, the heat treatment increased the roughness and hence strain in substrate-bound atomically thin BN nanosheets.

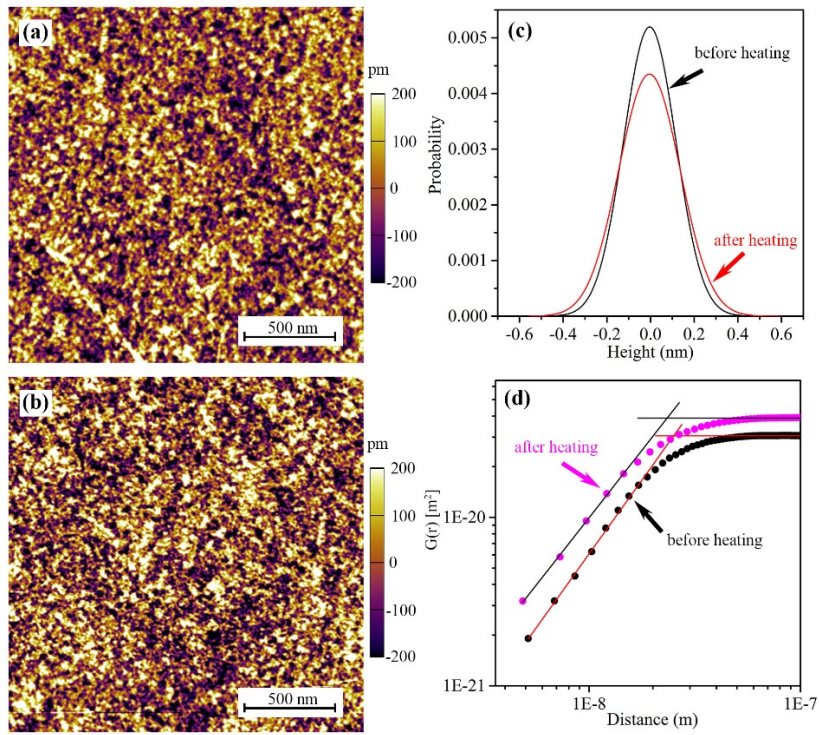

**Figure 3.** AFM images of a 1L BN on SiO₂/Si before (a) and after (b) heat treatment at 400 °C in Ar for 1 h; height distribution (c) and height-height correlation functions (d) of the nanosheet before and after the heat treatment.

As aforementioned, the increased roughness after heat treatment was caused by a mismatch of the TEC between SiO₂ substrate and BN nanosheets. The strain change introduced by the mismatch at certain temperature can be estimated by:[82]

$$\Delta\varepsilon = \int_{RT}^{T_m} \left( \alpha_{BN}(T) - \alpha_{SiO_2}(T) \right) dT \qquad (6)$$

where $\alpha_{SiO_2}$ and $\alpha_{BN}$ are the temperature-dependent TEC of SiO₂ and BN nanosheets, respectively. We used $0.75\times10^{-6}$/K for $\alpha_{SiO_2}$ and $-2.9\times10^{-6}$/K for $\alpha_{BN}$ in calculation.[49] Hence, our heat treatment at 400 °C should cause an increase of compressive strain of $-0.14\%$ in BN nanosheets. In other words, BN nanosheets expanded while SiO₂ contracted during cooling down stage, increasing the biaxial compressive strain in the nanosheets.

After the heat treatment, the Raman frequencies of substrate-bound 1-3L BN upshifted. The G band frequency of 1L BN increased from $1369.6\pm0.6$ cm⁻¹ to $1372.6\pm0.4$ cm⁻¹ after annealing, representing an upshift of $3.0\pm0.7$ cm⁻¹ (Figure 4b and c). The Raman upshifts observed from the heated 2L and 3L BN nanosheets are $2.7\pm0.8$ and $2.2\pm0.6$ cm⁻¹, respectively (Figure 4d and e). In contrast, no change of G band frequency was detected from bulk hBN before and after the same heat treatment (Figure 4a). It is clear that strain is the key factor that determines the Raman frequency of atomically thin BN nanosheets. We used the average Raman upshift of 1L BN after heat treatment to calculate the strain change:[29]

$$\varepsilon = -\Delta\omega_G / 2\gamma\omega_G^0 \qquad (7)$$

where $\Delta\omega_G$ is G band frequency shift, $\gamma$ is Grüneisen parameter of hBN (0.64 deduced from Figure 2e), $\omega_G^0$ is the G band frequency of unstrained BN (we used 1367 cm$^{-1}$). The strain change calculated based on the average upshift (3.0±0.7 cm$^{-1}$) of the 1L BN nanosheets after heat treatment is –0.17±0.04%. This value is basically in agreement with that estimated using Eq. 6.

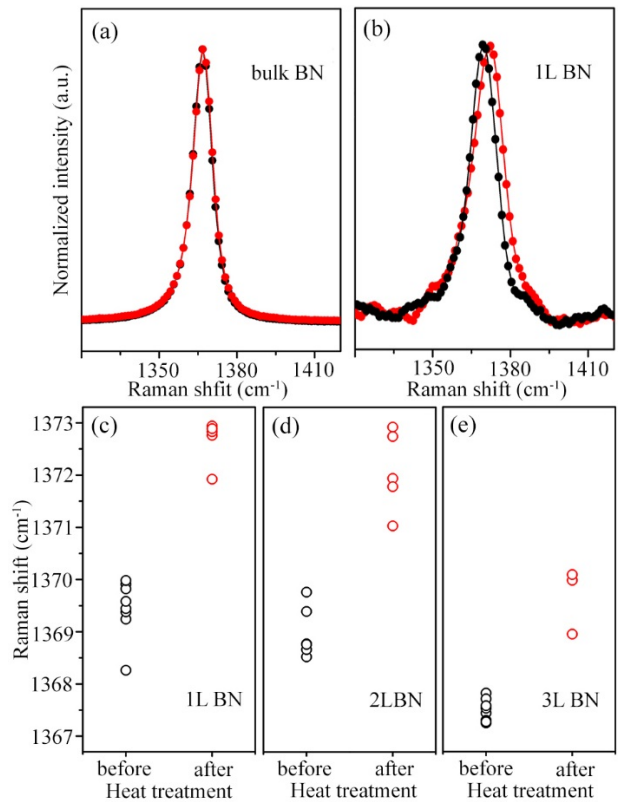

**Figure 4.** Raman spectra of bulk (a) and 1L (b) BN on SiO$_2$/Si, and Raman G band frequency shifts of 1L (c), 2L (d) and 3L BN (e) on SiO$_2$/Si before (black) and after (red) the heat treatment.

In summary, the Raman frequency of monolayer and few-layer BN measured from suspended nanosheets was similar to that of bulk hBN, suggesting that the $E_{2g}$ mode of BN did not depend sensibly on the thickness. This was justified by DFT calculations at PBE-opt-vdW and HSE06 levels of theory. Our simulations also indicated that the inclusion of exact exchange from the Hartree-Fock theory improved the accuracy of the calculated vibrational modes and gave remarkably improved agreement to the experimental data. In contrast, atomically thin BN on $SiO_2/Si$ showed upshifted Raman G bands with decreased thickness. This was due to higher flexibility of atomically thin nanosheets, which could follow the uneven surface of substrate more closely and hence gain more compressive strain. The substrate-induced strain in atomically thin BN was further increased by heating treatment, and as the result, further upshifts of the G band was observed.

**Experimental Section**

*Preparation of BN nanosheets.* The substrate-bound and suspended BN nanosheets were mechanically exfoliated from single-crystal hBN using Scotch tape on 90 nm $SiO_2/Si$ substrate with and without pre-fabricated 1.3 μm wells. An Olympus BX51 optical microscope equipped with a DP71 camera was used to locate atomically thin BN, and a Cypher AFM (Asylum Research) was employed to measure their thickness. The AFM images for estimating roughness of the nanosheets were taken with 512×512 pixels in contact mode. The Raman spectra of atomically thin and bulk BN were collected using a Renishaw inVia Raman microscope with a 514.5 nm laser. All Raman spectra were calibrated with the Raman band of Si at 520.5 cm$^{-1}$. An objective lens of 100x with a numerical aperture of 0.9 was used (*i.e.* laser spot size of ~0.9 μm). In order to further introduce strain to BN nanosheets on $SiO_2/Si$, the samples were heated in Ar at 400 °C for 1h.

*Theoretical calculation.* Theoretical calculations were performed using the DFT formalism as implemented in the Vienna *ab initio* simulation package (VASP).[83,84] For the calculations of phonon modes, both PHONON[85,86] and Phonopy[87] were used. A 2x2x1 supercell was used for these calculations. For the calculation of Raman band frequency, PHONON was used implementing the PEAD method. The optB88-vdW functional[88] was used along with a plane-wave cutoff of 800 eV combined with the projector-augmented wave (PAW) method.[89,90] Atoms were allowed to relax under the conjugate-gradient algorithm until the forces acting on the atoms were less than $1x10^{-8}$ eV/Å. The self-consistent field (SCF) convergence was also set to $1.0x10^{-8}$ eV. Relaxed lattice constants were found to be **a**=**b**=2.50976Å for the monolayer system and **a**=**b**=2.5110 Å for the bi and trilayer systems which is in excellent agreement with experiments. A 20 Å vacuum space was used to restrict interactions between images. A 12x12x1 gamma-centered **k**-grid was used to sample the Brillouin zone for all systems. K-sampling was increased to 24x24x1 and there was no appreciable difference seen in the values obtained.

AUTHOR INFORMATION

**Notes**

The authors declare no competing financial interests.

ACKNOWLEDGMENT

L.H.Li thanks the financial support from the Australian Research Council under Discovery Early Career Researcher Award (DE160100796). Y.Chen gratefully acknowledges the funding support from the Australian Research Council under the Discovery project. This work was performed in part at the Melbourne Centre for Nanofabrication (MCN) in the Victorian Node of the Australian National Fabrication Facility (ANFF). D.Scullion thanks the studentship from the EPSRC-DTP award. E.J.G.Santos acknowledges the use of computational resources from the UK national high-performance computing service, ARCHER, for which access was obtained via the UKCP consortium and funded by EPSRC grant ref EP/K013564/1; and the Extreme Science and Engineering Discovery Environment (XSEDE), supported by NSF grants number TG-DMR120049 and TG-DMR150017. The Queens Fellow Award through the start-up grant number M8407MPH and the Sustainable Energy PRP are also acknowledged.